\documentclass[11pt]{article}
\usepackage{amsmath}
\usepackage{graphicx}
\usepackage{amsfonts}
\usepackage{amssymb}
\usepackage{epsf}
\usepackage{latexsym}

\def\q{\underline{q}}

\def\d{\mbox{d}}

\def\sT{{\mbox{\scriptsize T}}}

\def\sfA{\mbox{\sffamily{\scriptsize A}}}
\def\sfa{\mbox{\sffamily{\scriptsize a}}}
\def\sfb{\mbox{\sffamily{\scriptsize b}}}
\def\sfc{\mbox{\sffamily{\scriptsize c}}}

\def\sL{\mbox{\sffamily L}} 
\def\sH{\mbox{\sffamily H}}

\def\md{\mbox{\scriptsize d}}

\def\sN{\mbox{\scriptsize N}}

\def\NSI{Na\"{\i}ve Schr\"{o}dinger Interpretation }
\def\CPI{Conditional Probabilities Interpretation }
\def\be{\begin{equation}}
\def\ee{\end{equation}}
\def\bea{\begin{eqnarray}}
\def\eea{\end{eqnarray}}
\def\fn{\footnote}
\def\pa{\partial}
\def\d{\textrm{d}}

\def\cr{\mbox{\scriptsize{\bf $\mbox{ } \times \mbox{ }$}}}
\def\hat{\widehat}
\def\R{\mbox{\underline{R}}}

\def\sL{\mbox{\scriptsize L}}
\def\sH{\mbox{\scriptsize H}}

\def\su{\mbox{\scriptsize universe}}

\def\ttH{\mbox{\tt{H}}}
\def\ttM{\mbox{\tt{M}}}
\def\ttD{\mbox{\tt{D}}}
\def\fE{\mbox{\sffamily E}}
\def\fH{\mbox{\sffamily H}}

\def\fL{{\cal L}}

\def\fP{\mbox{\sffamily P}}
\def\fQ{\mbox{\sffamily Q}}
\def\fR{\mbox{\sffamily R}}
\def\fS{\mbox{\sffamily S}}
\def\fT{\mbox{\sffamily T}}
\def\fU{\mbox{\sffamily U}}
\def\fV{\mbox{\sffamily V}}

\begin{document}

\begin{titlepage}

\begin{center}

\vspace{.3in}

{\Huge{\bf RECORDS THEORY}} 

\vspace{.3in}

{\large{\bf Edward Anderson}}$^1$

\vspace{.3in}

\noindent{\em\large Peterhouse, Cambridge, U.K., CB21RD;}

\vspace{.1in}

\noindent{\em\large DAMTP, Centre for Mathematical Sciences, Wilberforce Road, Cambridge, U.K., CB30WA.}

\vspace{.1in}

\noindent{\large Work started at {\em Department of Physics, Avadh Bhatia Laboratory, University of Alberta}}

\end{center}

\vspace{.3in}


\begin{abstract}

In quantum gravity, one seeks to combine quantum mechanics and general relativity.  
In attempting to do so, one comes across the `problem of time' impasse: 
the notion of time is conceptually different in each of these theories.  
In this paper, I consider the timeless records approach toward resolving this.  
Records are localized, information-containing subconfigurations of a single instant.
Records theory is the study of these and of how science (or history) is to be abstracted from 
correlations between them.  
I critically evaluate motivations for this approach that have previously appeared in the literature. 
I provide a ground-level structure for records theory and discuss what kind of further tools are needed, 
illustrated with some toy models: ordinary mechanics, relatonal particle dynamics, detector models and 
inhomogeneous perturbations about homogeneous cosmology.

\end{abstract}


\vspace{.3in}

\mbox{ }

%

\noindent{\bf PACS numbers 04.60-m, 04.60.Ds}

\mbox{ }

\vspace{3in}

\noindent$^1$ ea212@cam.ac.uk

\end{titlepage}

\section{Introduction}\label{Intro}

Although there are older notions of records in the philosophical literature (see e.g. 
Reichenbach \cite{Reichenbach} or Denbigh \cite{Denbigh}), 
this paper mostly concerns records in the modern physics literature, where they are linked to specific 
and partly technical Quantum Cosmology and Quantum Gravity issues 
(Sec \ref{POT}, \ref{rad}, \ref{QCos}).
Moreover, this literature on records is heterogeneous, splitting into the following three strands.  

\noindent 1) Bell \cite{Bell} and Barbour \cite{B94II, EOT}\fn{See Butterfield \cite{Butterfield} 
for a study of differences between these works.} reinterpret Mott's calculation \cite{Mott} of how 
$\alpha$-particle tracks form in a bubble chamber as a paradigm for Records Theory.  
Barbour, Halliwell \cite{HalliMott, Hallioverlap} and Castagnino--Laura \cite{CaLa, CastAsym} 
have argued for Quantum Cosmology to be studied analogously.   
Barbour's approach also involves reformulating classical physics in timeless terms 
\cite{BB82, B94I, RWR, Phil, Lan2}.  
It places emphasis on timelessness casting mystery \cite{EOT} upon why `ordinary physics' works,  
and on the configuration of the universe as a whole.     

\noindent 2) Page and Wootters \cite{CPI} and Page \cite{PAOT, Page} have put forward a {\it\CPI} for 
Quantum Cosmology  (see also the comments, criticisms and variants in \cite{K92, Kuchar99, Giddings, 
Pullin}), which improves on the {\it\NSI} \cite{NSI, HP86, HP88, UW89} (see Sec 4.5) in placing its 
emphasis on subconfigurations of the universe within a single instant.  
In this approach, ordinary physics of subconfigurations ends up being explainable in a familiar fashion 
through other subconfigurations providing a clock for them.      
While its conceptual basis is sound, it is not clear whether subsequent computations done in the 
literature follow uniquely from this conceptual basis, nor whether Page's later work involving  
memories will be amenable to mathematical implementation (See Sec 4.4).

\noindent 3) Gell-Mann--Hartle \cite{GMH} and Halliwell \cite{H99} have considered a `Records Theory 
within Histories Theory'.\fn{At the simplest level, a history is a Feynman path integral over an appropriate 
notion of time.} {\it Histories Theory} \cite{Hartle93} (Sec 2.1) is not primarily timeless [though the 
emphasis is on (sub)histories rather than (sub)configurations and times], but a Records Theory sits 
within it.   
Records Theory in this approach benefits by inheriting part of the structural framework that has been 
developed for Histories Theory.

\mbox{ }

\hspace{0.5in} Records are ``{\it somewhere in the universe that information is 
stored when} 

\hspace{1.25in} {\it histories decohere}" (p 3353 of \cite{GMH}).\hspace{2in} \mbox{(0)}   

\mbox{ }

This paper's Records Theory does not follow one of these strands but is rather a synthesis of 
elements drawn from each of them, to be subjected to testing using toy models, and refined if required.    
In outline, I consider {\it records} to be information-containing subconfigurations of a single instant.  
{\it Records Theory} is then the study of these and how dynamics (or history or science) is to be 
abstracted from correlations between same-instant records.  
It is to make this abstraction meaningful that I insist on records being subconfigurations rather than 
whole instants. 
In this way, one can get round some of Barbour's obstacles as regards why we appear to experience dynamics 
within a timeless universe -- it is an overall-timeless universe, but subsystems can provide approximate 
relative time standards for each other, and what is habitually observed is the dynamics of subsystems 
rather than of the whole universe \cite{BS, B94I, B94II, EOT, 06I, 06II, Pullin}.

For adopting a Records Theory approach to profitable, I argue that records should have the following 
properties.   
Records should be {\it useable}, in that 1) their whereabouts [c.f. (0)] should be spatially-localized 
subconfigurations of the universe, for whatever notion of space that one's theory has and restricted to 
the observationally accessible part thereof.  
2) They should also belong to a part of the subconfiguration space for which our imprecision of 
identification of the subconfiguration by how we observe it does not too greatly distort the extraction 
of information.  
Records should also be {\it useful}: their information content [c.f. (0)] should be high enough and of the 
right sort of quality to enable reliable measures of correlation to be computed.\fn{This idea expands 
on Denbigh's \cite{Denbigh} realization that records needn't all be orderly.  
Also note that, within approach 3), Halliwell's study \cite{H99} of imperfect records and simple  
detectors could be viewed as a first development of this notion of useful records.}  
As not all systems that one could be dealing with have instants solely of this nature, one would expect 
that a Records Theory approach would not always be profitable.

Barbour's conceptualization in particular (and also Page's) additionally require semblance of dynamics 
to emerge from the records.      
Barbour furthermore asks \cite{B94II, EOT} whether there are any {\it selection principles} for such 
records (which he calls `time capsules'; the bubble chamber with the $\alpha$-particle 
track within is a such).    
If these features are to be incorporated, one would additionally need a (relative) measure of semblance 
of dynamics.   
This would likely have some links with the notions of information content and quality, but would appear 
to require further input as a world could be detailed and nevertheless have no global Arrow of Time.
Barbour furthermore conjectures \cite{EOT} that the selection priciple is through the shape of the 
configuration space being such that the wavefunction is concentrated in regions containing `time 
capsules'.
However, he does not supply any evidence for this based on mathematical models.   
The {\it Semiclassical Approach} to Quantum Gravity (see Sec 2.1) might explain -- or supplant -- 
Barbour's proposal \cite{06I, 06II, SemiclI}, while {\it Branching Processes} and Histories Theory 
\cite{Hartle93} may provide alternative selection principles for records (see Sec 7.3 for more).    

\mbox{ }

In Sec \ref{Motivation} and Sec 4.4, I assess how Records Theory has been motivated in the literature. 
Records Theory has been motivated by 1) universality: it is a conceptual scheme for any kind of physical 
system (though it is not the only such conceptual scheme).      
2) By there not being limitations to conceptualizing about change, processes, dynamics, 
history and science in the timeless terms of Records Theory.  
I provide some evidence for this by looking at the question types that these conceptualizations can 
address and then finding ways of converting becoming questions into being questions.  
Some of these conversions, however, can be highly cumbersome in practise.    
3) Records Theory is furthermore likely to be useful in the study of closed systems, 
including Quantum Cosmology's study of the universe as a whole.  
4) It is also one natural perspective for Quantum Gravity: canonically quantizing General 
Relativity (GR) \cite{ADM, Dirac, Battelle, DeWitt} 
encounters serious conceptual and technical difficulties, among which one major conceptual problem is 
the Problem of Time (see e.g. \cite{Battelle, EarlyKuchar, Kuchar81, UW89, K92, Isham93, B94II, 
Kuchar99, Kiefer}).  
This problem's origin is in time playing incompatible roles in the standard formulations of GR 
and of QM (see e.g.  
\cite{Isham93, B94I, Kieferseminar}).   
Fundamental timelessness is one natural and clear resolution of this incompatibility,  
and Records Theory is one approach to timelessness.  
5) Records Theory has also been motivated in the literature by Canonical Quantum GR's 
Wheeler--DeWitt equation \cite{Battelle, DeWitt} giving rise to a frozen formalism.  
6) By `now' being what we experience.
7) By consideration of what is operationally meaningful.  
I provide arguments {\sl against} these last three.  
In the case of 7), I argue that the view that Histories Theory is primarily reconstruction 
from records is not a motivation but rather an assertion to be demonstrated, because the records present 
in nature may not in general be of sufficient quality to be able to reconstruct history.

I then propose in Sec 4 a ground-level structure for Records Theory which parallels some of the 
ground-level structure in Histories Theory.  
As it should not be discounted that it may be possible to set up a distinct and useful Records Theory 
without assuming all the levels of structure of Histories Theory, I do not for the moment demand any 
of History Theory's higher levels of structure.  
I then comment on the useability, usefulness and correlation aspects of records in Sec 4--6 (more 
technical detail of these will be provided elsewhere \cite{ARec2}).  
I illustrate Records Theory with the toy models of Sec \ref{Toys}: ordinary mechanics, 
relational particle mechanics\fn{These and their further use as toy models for GR are further discussed 
in \cite{BB82, Hartle88II, BS, K92, B94I, B94II, EOT, AConcat, ABFO, ABFKO, 06I, 06II, Phil, SemiclI, 
SemiclII, Triangle, FORD, 07II}.  
In particular, they share a number of features with the inhomogeneous perturbation model \cite{FORD}.}, 
detector models and inhomogeneous perturbations about homogeneous cosmology \cite{HallHaw, +IP} (a 
setting which is relevant as regards the origen of galaxies and of temperature variations in the CMB).

\section{Some motivations for Records Theory}\label{Motivation}

\subsection{The Problem of Time in Canonical Quantum General Relativity}\label{POT}

The canonical formulation of GR traditionally uses as its base objects the 3-metric variables 
$h_{\alpha\beta}(x^{\mu})$ on spatial slices $\Sigma$ of spacetime.\fn{
I use $( \mbox{ } )$ for functions, $[ \mbox{ } ]$ for functionals and $( \mbox{ } ; \mbox{ } ]$ with 
function arguments before the semicolon and functional arguments after it.  
I consider compact without boundary $\Sigma$ for simplicity (this is not restrictive).  
The determinant, covariant derivative and Ricci scalar associated with $h_{\alpha\beta}$ are denoted 
respectively by $h$, $\nabla_{\alpha}$ and ${\cal R} = {\cal R}(x^{\mu}; h_{\alpha\beta}]$.  
$\Lambda$ is the cosmological constant.  
With quantum cosmological applications in mind, I include a single minimally coupled scalar field 
$\phi$; $\pi_{\phi}$ is the momentum conjugate to $\phi$ and $\fV(\phi)$ is the scalar field potential.} 
%
The momenta conjugate to these are denoted by $\pi^{\alpha\beta}(x^{\mu})$ and are closely related to 
the extrinsic curvature of that space within spacetime.  
Canonical GR is a constrained theory.  
It is governed by 1) the linear {\it momentum constraint}
\be
{{\cal M}}^{\alpha}(x^{\mu}; h_{\alpha\beta}, \pi^{\alpha\beta}] \equiv  -2\nabla_{\beta}\pi^{\alpha\beta}  
+ \pi_{\phi}\pa^{\alpha}\phi = 0 
\label{GRmom} \mbox{ } ,
\ee
which is interpretable as the the physics being encoded within the 3-geometry rather than on the 
coordinate grid painted upon it.  
Hence this canonical formulation of GR is a {\it geometrodynamics}.  
2) The quadratic scalar 1-density {\it Hamiltonian constraint}\fn{Here
the DeWitt raising and 2 index to 1 index map `$v_{\tilde{A}} = v^{\alpha\beta}$' has been used.  
$N^{\tilde{A}\tilde{B}} = N_{\alpha\beta\gamma\delta} = \frac{1}{\sqrt{h}}
\left\{
h_{\alpha\gamma}h_{\beta\delta} - \frac{1}{2}h_{\alpha\beta}h_{\gamma\delta}
\right\}$ 
is the DeWitt supermetric (= inverse of GR's kinetic metric, $M_{\tilde{A}\tilde{B}}$), whose determinant 
is $M$.}
\be
{\cal H}(x^{\mu}; h_{\alpha\beta}, \pi^{\alpha\beta}] \equiv 
N^{\tilde{A}\tilde{B}}\pi_{\tilde{A}}\pi_{\tilde{B}} 
+ \frac{\pi_{\phi}^2}{2\sqrt{h}} - 
\sqrt{h} {\cal R} + \sqrt{h}2\Lambda + \sqrt{h}\fV(\phi) = 0 
\mbox{ } , 
\label{GRham}  
\ee
whose interpretation is more problematic.

At the quantum level, one then has the quantum momentum constraint  
\be
\hat{{\cal M}}^{\alpha}\Psi \equiv  - 2\nabla_{\beta} \frac{\delta }{\delta h_{\alpha\beta}(x)} \Psi +  
\pa^{\alpha}\phi\frac{\delta}{\delta\phi} \Psi = 0 
\label{GRQmom}
\ee
which is likewise interpretable as the wavefunction of the universe $\Psi$
depending on the 3-geometry rather than on any coordinate grid painted upon it.  
However, the quantum Hamiltonian constraint that one obtains -- the Wheeler--DeWitt equation (WDE) 
\cite{DeWitt, Battelle}, 
\be
\hat{\cal H}\Psi = 
\left\{
- 
{\hbar^2}
\left\{
`\frac{1}{\sqrt{M}}\frac{\delta}{\delta h^{\tilde{A}}}
\left\{
\sqrt{M}N^{\tilde{A}\tilde{B}}\frac{\delta}{\delta h^{\tilde{B}}}
\right\}\mbox{'}
+
\frac{1}{2\sqrt{h}}\frac{\delta^2}{\delta\phi^2}
\right\}
-  
{\sqrt{h}{\cal R}} + \sqrt{h}\fV(\phi)
\right\}  
\Psi 
+ {\sqrt{h}2\Lambda   }\Psi = 0
\label{WDE} \mbox{ } ,
\ee 
appears to be\fn{The inverted commas denote that the WDE has additional technical problems: there are 
operator-ordering ambiguities (the ordering I give here is the Laplacian one, see e.g. \cite{HP86, 07II} 
for motivation) and regularization is required, while what functional differential equations mean 
mathematically is open to question.} 
a timeless equation: a {\it stationary}, or {\it time-independent Schr\"{o}dinger equation} (TISE)).
This {\it frozen formalism} that the WDE seems to imply can be traced back to the diffeomorphism 
invariance of GR, with its absense of any appended external parameter such as that which conventionally 
plays the role of time in Newtonian theory.  
The frozen formalism is a prominent part of the Problem of Time (POT) for Canonical Quantum GR (though 
this is by no means the only manifestation of the POT nor do all possible approaches to Canonical 
Quantum GR, let alone Quantum Gravity, involve such an equation, see especially \cite{K92, Isham93}).

Various interpretational strategies for dealing with the POT are discussed in  
\cite{DeWitt, EarlyKuchar, Kuchar81, UW89, K92, Isham93, Kuchar99, EOT, Kiefer, 06II},  
which are references of use throughout this subsection.  
Some of the differences in strategy result from two prevalent and conflicting philosophical positions 
concerning time \cite{Denbigh}: A) that time is fundamental, or 
B) that time should be eliminated from one's conceptualization of the world.  
These two perspectives can be related in various ways to schemes for the world and the scientific 
enterprise therein which question-types involving `becoming' fundamentally make sense, or only 
those involving `being'.\fn{Hawking, 
Page, Wootters and Barbour \cite{NSI, HP86, HP88, CPI, PAOT, Page, B94I, B94II, EOT, RWR} have written 
in favour of being from which the semblance of becoming can arise    
[although not much quantitative progress has been made with the semblance part].  
Kucha\v{r} favours becoming, both in his research and in his review \cite{K92}, while Isham's review 
\cite{Isham93} is more conciliatory. 
%
Hartle and Halliwell have considered both \cite{Hartle93, GMH, H99, HT, Hallioverlap}.  
I argue that one should give a fair hearing to each strategy from whichever perspective is appropriate to 
it. 
This paper principally investigates timeless strategies, for which the appropriate perspective is being.} 

\mbox{ }

The following strategies have been put forward for addressing the POT.  

\noindent 1) Internal Time: There might be a fundamental classical time in all circumstances. 
This might not be obvious through its being `hidden within' the theory.  
It would be found by replacing ${\cal H} = 0$ by its classical solution for a momentum variable that 
is perhaps new (obtained by a canonical transformation): $P_{  {\cal T}_{\mbox{\tiny int}}  }(x) = 
P_{  {\cal T}_{\mbox{\tiny int}}  }(x; {\cal T}(x), Q^{\sT}(x), P^{\sT}(x)] \equiv 
{\cal H}^{\mbox{\scriptsize true}}(x; {\cal T}_{\mbox{\scriptsize int}}, Q_{\sT}, P^{\sT})$.\fn{This 
is the simpler `time function' version; there is also a 4-component `embedding variable' version. 
T indexes the true dynamical degrees of freedom.} 
Then quantization gives a time-{\sl dependent} Schr\"{o}dinger equation (TDSE) 

\noindent $i\hbar\frac{\delta \Psi}{\delta {\cal T}_{\mbox{\tiny int}}} =  
\widehat{{\cal H}}^{\mbox{\scriptsize true}}
(x; {\cal T}_{\mbox{\scriptsize int}}, Q_{\sT}, \widehat{P}^{\sT})$
that supplants (\ref{WDE}).      
An example of hidden internal time candidate is GR's York time \cite{York72, K92, Isham93, Kuchar81} 
which is proportional to the constant mean curvature slices of spacetime, 
$h_{\mu\nu}\pi^{\mu\nu}/\sqrt{h}$ = Const, that generalize the maximal slices, $h_{\mu\nu}\pi^{\mu\nu} = 0$.

\noindent 2) Emergent Time: in certain, possibly predominant, circumstances, an {\it emergent notion of 
time} may occur.      
One possibility for this is that in situations in which the Born--Oppenheimer approximation 
$\Psi = \psi(\mbox{H})|\chi(\mbox{H}, \mbox{L}\rangle$ for H `heavy' and L `light' degrees of freedom 
and the WKB approximation $\psi(\mbox{H}) = e^{iW(\mbox{\scriptsize H})/\hbar}$ are applicable, an 
emergent time drops out of the WDE \cite{DeWitt, HallHaw, Kiefer, SemiclI, SemiclII}.  
For, $\hbar^2N^{\tilde{A}\tilde{B}}\frac{\delta^2\Psi}{\delta h^{\tilde{A}}\delta h^{\tilde{B}}}$ 
contains $\hbar^2N^{\tilde{A}\tilde{B}} \frac{i}{\hbar}
\frac{\delta W}{\delta h^{\tilde{A}}} \frac{\delta|\chi\rangle}{\delta h^{\tilde{B}}} = 
i\hbar N^{\tilde{A}\tilde{B}}\pi_{\tilde{A}}\frac{\delta|\chi\rangle}{\delta h^{\tilde{B}}}$ 
by the Hamilton--Jacobi relation for the momentum, and this expression contains 
$i\hbar \frac{\delta h^{\tilde{B}}}{\delta (t = t^{\mbox{\tiny WKB}})}
\frac{\delta|\chi\rangle}{\delta h^{\tilde{B}}}$ by the momentum--velocity relation, which is 
$i\hbar\frac{\delta|\chi\rangle}{\delta t^{\mbox{\tiny WKB}}}$ by the chain-rule so that one has a 
TDSE for the light degrees of freedom with respect to a time standard that is (approximately) provided 
by the heavy degrees of freedom.  
An issue here is that (semi)classical conditions need not always occur -- guarantee of a classical 
`large' as in the Copenhagen Interpretation of QM has been cast aside in Quantum Cosmology and will 
then by no means be recovered in all possible situations.

\noindent 3) Timelessness: at face value, (\ref{WDE}) is suggestive of there fundamentally being no time 
for the universe as a whole in spatially compact without boundary GR.  
These strategies aim to supplant `becoming' with `being' at the primary level \cite{Page, B94II, EOT, 
GMH, H99, HT, HalliDodd, Hallioverlap}.    
History or dynamics are to be {\sl apparent notions} to be constructed from the instant  
\cite{GMH, H99, HalliDodd, Hallioverlap}. 
Examples of timeless strategies include the Na\"{\i}ve Schr\"{o}dinger Interpretation \cite{NSI, HP86, 
HP88, UW89} (`na\"{\i}ve' because it uses $\int\Psi_1^*\Psi_2 \d\Omega$ as its inner product with no 
heed for the constraints) and the wider-ranging Conditional Probabilities Interpretation \cite{CPI} 
which has been suggested to be general enough to cover all types of questions that occur in science 
\cite{Page, B94II, EOT}. 
(`Interpretation' here is meant in the sense of `interpretation of quantum theory'.)

\noindent 4) Histories Theory: instead, it could be the histories themselves which are primary, 
and the records that are constructs (see e.g. \cite{Hartle88II, GMH, Hartle93, +Hartle, H99}).  
[Then it is not clear whether Records Theory within History Theory is necessarily the same as Records 
Theory from first principles.]

Each of these strategies has problems if examined in detail.    
POT resolutions are usually taken to be required to work for a full theory of gravitation, so that 
establishing that they work for special cases or toy models is not enough.  
I in no way claim that the present paper is exhaustive as regards what problems timeless strategies 
have -- see \cite{K92, Isham93, Hartle93, Kuchar99} for plenty more issues.

\subsection{Records Theory as a radical solution of the POT?}\label{rad}

It has been argued \cite{B94I, B94II, EOT} that  
a records approach is natural if one takes the frozen formalism seriously. 
It has also often been argued (see e.g. \cite{EOT, Brout, HalliMott})  
that successors to the WDE through GR being supplanted would likely also have a frozen formalism, 
but I note that this is not always the case, e.g. there is a higher derivative theory for which one 
of the natural variable sets contains an already-explicit internal time \cite{BoulwareHorowitz}. 
One might then also question whether one should go beyond Kucha\v{r} and Isham's adherence to GR 
\cite{K92, Isham93} in investigating the POT.

Barbour \cite{B94I, B94II, EOT} and Barbour and Smolin \cite{BS} additionally suggested timeless records 
approaches as radical approachs that are attractive due to other, more conventional strategies' failures 
to resolve the POT.   
However, arguing by elimination is dangerous, firstly because we are unlikely to ever know what all the 
conceptual and technical options are.  
Secondly, I point out that the claim that the radical records approach has no (or fewer) problems is 
itself dubious, both because knowing few problems may merely reflect that it has been studied less than other 
approaches and because there {\sl are} salient problems as regards the recovery of 
a semblance of dynamics or history for subsystems: I provide some in Sec 7.

An alternative to thinking in terms of radical solutions is to view the frozen formalism not as a truth 
to be taken seriously but as a `dynamics observed within a timeless universe' `paradox' to be resolved. 
The Semiclassical Approach, Internal Time Approach and Histories Theory do look like having `paradox' 
resolving capacities (see e.g. Sec 2.1).  
Resolving it may well hinge on the difference between physics of a closed universe and physics in 
a closed universe.  
If a small subsystem is studied, there's little trouble with the larger and more influential exterior 
providing a background time.  
By this device, an essentially standard perspective on ordinary physics (such as in \cite{EOT}) or 
smaller scale cosmology (such as isolated gravitational perturbations) suffices.
Barbour's `end of time' idea \cite{EOT} that the timelessness mystery pervades ordinary physics can 
thus be undone.
%
%
Mysteries that may need radical addressing remain only for more obviously problematic cases that  
concern genuinely closed systems such as the whole universe.

\subsection{Records to be primary because `now is what we experience'?}\label{now}

Does the common intuition that `now is what we experience' make Records Theory a natural, reasonable 
approach to theoretical physics?  
However the perceived now is, from a becoming perspective, actually a `specious present' i.e. an 
integrated experience whose formation requires a timescale that is vastly greater than that of 
already-known quantum processes.  
This may undermine taking `now is what we experience' as a first principle in theoretical physics.

\subsection{The records perspective might help with Quantum Cosmology?}\label{QCos}

A primary practical motivation for studying such as the POT, Records Theory and Histories Theory is 
that Quantum Cosmology \cite{DeWitt, Misner} is not yet sufficiently developed at a conceptual level 
to provide a rigorous framework for inflationary scenarios (see e.g. \cite{HT, Hallioverlap}), 
which themselves have been argued to be useful not only through their various theoretical advantages 
over the plain Big Bang scenario but also through their good prediction of what the most recent CMB data 
should look like \cite{Inflation}.
It is via the above contact with testable assertions that hitherto philosophical contentions about QM (in 
particular as applied to closed systems such as the whole universe) may enter into mainstream physics.     
Suggestions as to how one might approach such conceptual issues in Quantum Cosmology include 
\cite{Bell, GMH, Hartle93, +Hartle, H99, HalliDodd, HT, Hallioverlap, B94II, EOT,  
HP86, HP88, UW89, CPI, K92, Isham93, Kuchar99, Page, Kiefer, +closed}.

How would a Records Theory perspective help with some of these issues?

\noindent 0) One might consider such as CMB inhomogeneities or the pattern and spectra of galaxies to 
involve useful records.  

\noindent 1) Within a histories perspective, the decoherence process makes records, but information is 
in general lost in the making.  
E.g. Halliwell \cite{H99} has shown that mixed states necessarily produce imperfect records.   
Furthermore, finding out where in the universe the information resides (i.e. where the records are) 
should be capable of resolving whether gravity decoheres matter or vice versa.\fn{Which 
occurs may depend on a situation-by-situation basis.}
Finally, decoherence is habitually linked with the emergence of (semi)classicality, so there may well be 
some bridge between Records Theory and the Semiclassical Approach.  

\noindent 2) In Barbour's approach, the GR universe is a collection of timeless records.  
One could then propose the Semiclassical Approach as providing the mechanism by which this description 
can nevertheless give a semblance of dynamics.  
A further idea of Barbour's (not yet backed by any evidence from mathematical models) is that the 
Arrow of Time might emerge {\sl from} records within a scheme that does not presuppose history.

\subsection{Records as a universal theoretical scheme}\label{uni}

As regards us probably not knowing in any detail which theory is correct in the Quantum Gravity 
regime, note that histories and records frameworks in general are not theory-sensitive.  
This should be contrasted with how some particular internal time candidates are tied to particular 
(families of) theories.  
The Semiclassical Approach's position is likely intermediate in this respect -- one would hope that the 
possibility of such an approximate regime is widespread among Quantum Gravity theories, albeit the rigorous 
justification of such a regime may depend on certain features not shared by all theories 
(as speculated in \cite{Isham93}).  
On the other hand, if a theory has a useable notion of internal time, arranging that theory into a 
Records Theory form might be viewed as a formalism to be used for only some applications.

\subsection{Records are what is operationally meaningful?}\label{op}

The most well-known old arguments about the physics being in the correlations  
are specifically not referring to Records Theory. 
E.g. what Wigner said on the subject is, in detail, (\cite{Wigner} p 145) ``{\it quantum mechanics only 
furnishes us with correlations between subsequent observations}", and Wheeler agreed with this 
(\cite{Battelle} p 295).  
Were one to wish to draw motivation from these statements, they would need to be rephrased 
and the original context would be lost.

That records are more operationally meaningful than the histories of the other universal scheme 
has been argued along the following lines.  
The study of records is how one does science (and history) in practise, whether or not one ascribes 
reality to whatever secondary frameworks one reconstructs from this (like histories, spacetimes or the 
local semblance of dynamics).  
Unfortunately, {\sl as motivation}, more careful inspection reveals this to be vague wordplay.  
This is because of the difference between the notion of records as in (some of the sub)configurations 
that the system provides and the notion of records as in things which are localized, accessible and of 
significant information content.  
As effective reconstruction of history requires the subconfigurations in question to have these in 
general unestablished properties, it is a {\sl question to address} rather than a preliminary motivation 
to have that records are more operationally meaningful than histories through possessing good enough 
qualities to permit a meaningful such reconstruction.    
Thus what one should do is 1) pin down where the ``somewhere" in (0) is (the central motivation in some 
of Halliwell's papers \cite{H99, HalliDodd}).    
2) Determine whether the record thereat is useful --  Gell-Mann--Hartle assert that what they call 
records ``{\it may not represent records in the usual sense of being constructed from quasiclassical 
variables accessible to us}" (p 3353 of \cite{GMH}).  
Also, it may be that the $\alpha$-particle track in the bubble chamber is atypical in its neatness and 
localization.  
For, bubble chambers are carefully selected environments for revealing tracks -- much human trial and 
error has gone into finding a piece of apparatus that does just that.  
$\alpha$-tracks being useful records could then hinge on this careful pre-selection, 
records in general then being expected to be poorer, perhaps far poorer, as suggested e.g. by the 
Joos--Zeh paradigm of a dust particle decohering due to the microwave background photons.  
In this situation, records are exceedingly diffuse as the information is spread around by the CMB 
photons.  
See Sec 5 and \cite{ARec2} for more details.

\section{Some toy models for illustrating Records Theory}\label{Toys}

Given the substantial technical difficulties of each Canonical Quantum GR strategy, it is worth 
considering conceptual questions for {\sl toy models}.

\subsection{Ordinary (conservative) mechanics}\label{Mech}

This has, as a simpler analogue of (\ref{GRham}), a quadratic energy constraint
\be
\ttH(q_{i\alpha}, p^{i\alpha}) \equiv n_{i\alpha j\beta}p^{i\alpha}p^{j\beta}/2 + \fV(q^{i\alpha}) - \fE 
= 0  
\label{EnO}
\ee
for $\fE$ a constant energy and $n_{i\alpha j\beta}$ the inverse constant diagonal mass matrix 
$\delta_{i\alpha j\beta}/m_i$.   
Then at the quantum level one has a TISE analogue of the WDE,\fn{For 
this example and for (\ref{TISE}), there is no ordering ambiguity so the simplest presentation 
(here used) is equivalent to the Laplacian one.}  
\be
\ttH\Psi = -\frac{\hbar^2}{2}n_{i\alpha j\beta}\frac{\pa}{\pa q_{i\alpha}}\frac{\pa}{\pa q_{j\beta}}\Phi + 
\fV(q^{i\alpha})\Phi - \fE\Phi = 0 \mbox{ } .  
\label{TISEO}
\ee

\subsection{Relational Particle Mechanics}\label{RPM}

There are other mechanics that share more features with GR.  
Consider the relational particle models (RPM's) in relative Jacobi coordinates\fn{Being 
relative coordinates, these automatically take care of the zero momentum constraint of the earlier 
relational particle model literature.  
Relative {\it Jacobi coordinates} $R_{i\alpha}$ \cite{Marchal} are chosen for their property of 
diagonalizing the kinetic term.  
The i-indices label relative coordinates; the summation convention does not apply to these indices.  
In relation to conventional position coordinates $q_{I\alpha}$ $I = 1$ to N with masses $m_I$,  
the relative Jacobi coordinates are interparticle (cluster) separations from $i = 1$ to n = N -- 1 
associated with (cluster) reduced masses $M_i$; I do not provide 
$M_i = M_i(m_J)$ relations here as these in any case depend on normalization convention, and 
Jacobi coordinates are nonunique for N $> 3$.  
$N_{i\alpha j\beta}$ is the inverse Jacobi mass matrix, $\delta_{ij}\delta_{\alpha\beta}/M_i$
I use d to denote the spatial dimension of a particle model.
$\fV$ is the potential energy of the model and $\fE$ is the total energy, which I consider to be a 
prescribed $\fE_{\su}$ rather than an a priori free parameter to be restricted to lie 
on a constructed eigenspectrum.  
The $P_{i\alpha}$ are the momenta conjugate to the $R_{i\alpha}$.}   
$\R_{i\alpha}$.  
These resemble GR in having 1) a quadratic energy constraint
\be
{\ttH} \equiv 
N_{i\alpha j\beta}P^{i\alpha}P^{j\beta}/2     
+ \fV(|R_{i\alpha}|) - \fE = 0 \mbox{ } 
\label{E}
\ee 
and 2) a linear constraint, the zero angular momentum constraint,  
\be
\ttM_{\alpha} \equiv {\epsilon_{\alpha}}^{\beta\gamma}R_{i\beta}P_{i\gamma} = 0 \mbox{ } .
\label{ZAM}
\ee
The former is interpretable as a timeless equation and the latter as the physics being encoded in the 
relative separations and angles (`relational quantities') rather than in any absolute angles.  
Scale-invariant RPM's additionally have a linear zero dilational constraint 
\be
\ttD  \equiv R_{i\alpha}P^{i\alpha} = 0 
\label{ZDM}
\ee
that is analogous to the maximal slicing condition in GR.  
In this case $\fV$ is solely a function of ratios of $|R_{i\alpha}|$.

Then at the quantum level, one has the quantum energy constraint 

\noindent
\be
\hat{\ttH}\Phi \equiv 
-\frac{\hbar^2}{2}N_{i\alpha j\beta}\frac{\pa}{\pa R_{i\alpha}   }
                                    \frac{\pa}{\pa R_{i\beta}    }\Phi +  
\fV(|R_{i\alpha}|)\Phi - \fE\Phi = 0 \mbox{ } ,
\label{TISE}
\ee 
which is another TISE analogue of the GR WDE.    
One also obtains the quantum zero angular momentum constraint
\be
\hat{\ttM}_{\alpha}\Phi \equiv  
{\epsilon_{\alpha}}^{\beta\gamma}R_{i\beta}\frac{\pa}{\pa R_{i\gamma}}\Phi = 0
\label{QMZAM}
\ee
i.e. that the wavefunction of the toy universe $\Phi$ depending on relational quantities rather than on 
any absolute angles, and, in the scale-invariant case, the quantum zero dilational momentum constraint 
\be
\hat{\ttD}\Phi  \equiv R_{i\alpha}\frac{\pa}{\pa R_{i\alpha}}\Phi = 0 \mbox{ } .  
\ee

Full reductions are available for 2d RPM's (of which the scale-invariant one is better behaved) 
allowing us to do quite a lot more with these particular models.  \cite{Kendall, Triangle, FORD}. 
Then one has a quadratic energy constraint 
\be
{\ttH}^{\mbox{\scriptsize red}} = \frac{1}{2}{\cal N}^{\sfa\sfb}({\cal Z}^{\sfc}){\cal P}_{\sfa}{\cal P}_{\sfb} + 
\fV(|{\cal Z}^{\sfc}|) - \fE = 0
\ee
for ${\cal Z}^{\sfc}$ the inhomogeneous coordinates (independent set of ratios of 
$z_{\sfA} = R_{\sfA}\mbox{exp}({i\Theta_{\sfA}})$ for $R_{\sfA}$, $\Theta_{\sfA}$ the polar Jacobi 
coordinates) and ${\cal N}^{\sfa\sfb}$ the inverse of the kinetic metric ${\cal M}_{\sfa\sfb}$, 
which is the Fubini--Study metric corresponding to the line element 
\be
\d D^2 = \frac{\{1 + \sum_{\sfc}|{\cal Z}_{\sfc}|^2\} \sum_{\sfc}|\d {\cal Z}_{\sfc}|^2 - 
                 |\sum_{\sfc} {\cal Z}_{\sfc} \d \overline{{\cal Z}}_{\sfc}|^2}
              {\{1 + \sum_{\sfc}|{\cal Z}_{\sfc}|^2\}^2}  \mbox{ } 
\label{FS} 
\ee
where the overline denotes complex conjugate and $|\mbox{ }|$ the complex modulus.  
Then at the quantum level and employing the Laplacian ordering, one obtains a TISE analogue 
of the WDE,
\be
\hat{\ttH}^{\mbox{\scriptsize red}}\Phi = \frac{1}{2}\frac{1}{\sqrt{\cal M}}\frac{\pa}{\pa{\cal Z}^{\sfa}} 
\left\{
\sqrt{{\cal M}}{\cal N}^{\sfa\sfb}\frac{\pa}{\pa{\cal Z}^{\sfb}}
\right\}
\Phi + \fV(|{\cal Z}^{\sfc}|)\Phi - \fE\Phi = 0 \mbox{ } .   
\label{TISE2d}
\ee

\subsection{Detector models}\label{Detector}

Following Halliwell \cite{H99}, one could couple such as an up--down detector to one's mechanics model, 
or include a harmonic oscillator detector within one's mechanics model.  
This can hold information about one Fourier mode in the signal, thus exemplifying that even very simple 
systems can make imperfect records.

\subsection{Inhomogeneous perturbations about homogeneous cosmologies}\label{HH}

What I consider is the linearized theory of perturbations about a homogeneous cosmology but not the WKB 
approximation, i.e. equations from early on in Halliwell--Hawking \cite{HallHaw} rather than their full 
development as a semiclassical scheme.  
I do not supply the form of the equations for this model here since they are complicated, but comment 
that they consist of a WDE and a nontrivial linear momentum constraint.  
It should also be noted that this model is no longer a mere toy, at least for some purposes such as studying 
the origin of galaxies or of microwave background perturbations.

One virtue of the RPM's \cite{FORD} is to have a number of features in common with this more complicated 
situation.  
E.g. they both have nontrivial linear constraints (momentum constraint versus zero total angular 
momentum constraint).  
Another similarity is explained in Sec 4.1.

\section{Configuration space structure and useable records}\label{GroundLevel}

\subsection{First level of structure}

For the moment, I work at the classical level.\fn{I comment on the QM counterparts in Sec 4.5.  
The classical concepts and tools considered here are significant at the quantum level, whether by 
themselves or as structures on which to base quantum concepts and tools.}     
%
%
A {\it configuration} $Q_{\Delta}(p)$ is a set of particle positions and/or 
field values, multi-indexed by the set $\Delta$ which covers both particle/field species labels and 
whatever `tensorial' indices each of these may carry.  
The present use of $p$ is as a fixed label.    
If $\Delta$ multi-indexes the whole of a universe model's contents, I denote it by $\Upsilon$;  
$Q_{\Upsilon}(p)$ is a {\it universal instant}.


Hierarchical, nonunique splittings into subsystems can then be construed: 
$Q_{\Gamma}(p)$ is a subsystem of $Q_{\Delta}(p)$ if $\Gamma$ is a subset of the indexing set $\Delta$.  
The {\it finest} such subdivision is into individual degrees of freedom; this is generally nonunique 
(e.g. under coordinate redefinitions). 
Such splits include: 1) {\it `Heavy--light' (H-L) splits} by which some $Q_{\sL}(p)$ are  
`more negligible' than the other $Q_{\sH}(p)$ for H, L a partition of $\Upsilon$.  
Then in the Semiclassical Approach, $Q_{\sH}(p)$ plays the role of approximate universal configurational 
instant that provides the approximate time standard for the L-physics (Sec 2.1).  
\noindent 2) Splits into the operationally-defined `studied subsystem' and the remaining 
`environment/background'.    
\noindent Note that 1) and 2) need not necessarily be aligned --   
one concerns what dominates the physics and the other concerns which part of the whole one beholds.


Two question-types that may be considered at this level are: 

\noindent Be$1^{\prime}$), does $Q_{\Delta}(p)$ have acceptable properties? (That covers 
both mathematical consistency and physical reasonableness).  

\noindent Be$2^{\prime}$) If properties of $Q_{\Delta}(p)$ are known, does this permit deduction of 
any observable properties of some $Q_{\Delta^{\prime}}(p)$ for $\Delta^{\prime}$ disjoint from $\Delta$? 
In other words, are there observable correlations between subconfigurations of a single instant?

\noindent E.g. within a RPM universe, if one's observed subsystem has angular momentum $L_0$, 
one can predict that the angular momentum for the rest of the universe is $-L_0$.  
[Verification of this by observations may not however be straightforward if the matter distribution 
for the rest of the universe is not local, or even possible if some of it is in an unobservable place.]


Many notions and constructions that theoretical physicists use (see e.g. \cite{Isham93}) 
%
%
additionally require consideration of sets of instants.  
A {\it configuration space of instants} is  $\fQ_{\Delta} = \{ Q_{\Delta}(p) \mbox{ } : \mbox{ } p$ 
a label running over a (generally stratified) manifold$\}$.  
This type of space is one example of {\it heap} -- a collection of instants -- other examples of which 
are considered in Sec 4.2. 
One defines subconfiguration spaces similarly.
Decomposition into subsystems is now a break-down into subspaces.

As examples of configurations which make up configuration spaces, 
an ordinary (absolute) particle mechanics configuration space is the set of possible positions of N 
particles, $\fQ$(N, d) = $\{\q_I, \mbox{ } I = 1$ to  N$\}$.  
Relative configuration space is the set of possible relative positions of N particles,  
$\fR$(N, d) = $\{\R_i, \mbox{ }  i = 1$ to $n = N - 1\}$. 
Relational configuration space is the set of possible relative separations and relative 
angles between particles ${\cal R}$(N, d) = $\{ \bar{R}_{\bar{\mbox{\i}}} = 1$ to dN -- d(d+1)/2 $\}$.  
Preshape space $\fP$(N, d) is the set of possible scale-free relative particle positions (nd -- 1 
independent ratios of relative coordinates).    
Shape space $\fS$(N, d) is the set of possible scale-free relational configurations (nd -- 2 
independent ratios of relative coordinates, which in 2d is parametrizable by the ${\cal Z}_{\sfa}$).
As regards geometrodynamics, a rather redundant configuration space is Riem($\Sigma$), the space of 
positive-definite 3-metrics $h_{\mu\nu}(x_{\gamma})$ on the 3-space of fixed topology $\Sigma$.  
A less redundant one is superspace($\Sigma$) = Riem($\Sigma$)/Diff($\Sigma$) (see e.g. \cite{Battelle}) 
and an even less redundant one is (something like) conformal superspace($\Sigma$) = Riem($\Sigma$)/
Diff($\Sigma$) $ \times $ Conf($\Sigma$) (see e.g. \cite{ABFKO}), for Diff($\Sigma$) the diffeomorphisms 
of $\Sigma$ and Conf($\Sigma$) the conformal transformations of $\Sigma$.        
For homogeneous cosmologies, the above spaces are all finite.  
E.g. Superspace becomes the well-known {\it minisuperspace}.  
Considering small inhomogeneous perturbations about this amounts once again to considering (simpler) 
infinite-dimensional spaces.   

\mbox{ }

While each $\fQ_{\Delta}$ corresponds to a given model with a fixed list of contents, one may not know 
which model a given (e.g. observed) (sub)configuration belongs to, or the theory may admit operations 
that alter the list of contents of the universe.    
Then one has a collection (grand heap) of (sub)configuration spaces of instants, $\fQ_{\Delta(\beta)}$, 
where $\beta$ parametrizes the collection.
For example, use 1) $\bigcup_{\sN \in \mbox{ \scriptsize $\mathbb{N}_0$ }}\fQ$(N, d) for 
a mechanics theory that allows for particle coalescence/splitting or creation/annihilation.
2) $\bigcup_{\Sigma \mbox{ \scriptsize compact without boundary}}$ superspace($\Sigma$) 
for a formulation of GR that allows for spatial topology change.  
From now on, I use $\fQ$ to denote a general (sub)configuration space or collection thereof.


Once one has a notion of configuration space $\fQ$, one can additionally have a second type of nonunique 
hierarchical splitting: {\it grainings} (i.e. the various ways that $\fQ$ can be partitioned).   
At least some grainings are definable without appeal to additional structures:  
the partition into grainings could be a classification by intrinsic properties of each $Q_{\Delta}(p)$.  
Graining defines a partial order $\prec$ on the subsets of $\fQ$.  
$A \prec B$ is termed `$A$ is finer grained than $B$', while 
$C \succ D$ is termed `$C$ is coarser-grained than `$D$'.  
The coarsest grained set is $\fQ$ itself, while the finest grained sets are each  
individual $q(p)$ (the points which make up the manifold $\fQ$).


As regards {\sl localization in space}, for example in RPM's one could consider the 3-body configurations 
in which the separation $s$ between particles 1, 2 is less than 1 km (an externally defined notion), 
or those in which in which particle 3 is within the sweep of $s$ (an internally defined notion).


As regards having a {\sl notion of closeness on configuration spaces} (or even on collections of them), 
sometimes one can do this by augmenting the configuration space to be equipped with a norm.  
E.g. on $\fQ$(N, d), these are the obvious unweighted, mass-weighted and inverse mass-weighted 
$\mathbb{R}^{\sN\md}$ norms; these may be considered to continue to play a role in more reduced 
configuration spaces through these inheriting structures that can be taken to be induced from them, such 
as the $\mathbb{R}^{nd}$ norm in $\fR$(N, d) and the chordal norm in $\fP$(N, d).\fn{For 
all that such spaces are termed relational, they still bear imprints of the absolute.  
Other examples are a residual sense of dimensionality and some topological aspects being inherited  
\cite{FORD}.}  
%
If the configuration space has a natural metric more complicated than the Euclidean one, one 
might be able to extend the above notion to the norm corresponding to that.  
For example, one could use the Fubini--Study norm on $\fS$(N, 2),  
while on Riem(M) one can to some extent use the inverse DeWitt line element (one problem being that 
this is indefinite). 

Another way is to intrinsically compute on each configuration a finite number of quantities, 
i: $\fQ \longrightarrow \mathbb{R}^n$, and then use the $\mathbb{R}^n$ norm 

\noindent
\be
D_{\mbox{\scriptsize Eucl}}^{\mbox{\scriptsize i}}(Q_{\Delta}, Q_{\Delta}^{\prime}) = 
||\mbox{i}(Q_{\Delta}) - \mbox{i}(Q_{\Delta}^{\prime})||^2
\ee
(though this is limited for some purposes by i having a nontrivial kernel).
For example, one can compare subconfigurations in $\fQ$(N, d), $\fR$(N, d) or ${\cal R}$(N, d) by 
letting i be the total moment of inertia for each subconfiguration, which is a mass-weighted norm.  
One could use the $\epsilon$ notion of almost alignment for each grouping of three 
particles: they are $\epsilon$ aligned if the largest angle in the triangle formed by their positions 
is $\geq \pi - \epsilon$ \cite{Kendall80}.  
In geometrodynamical theories, one could additionally compute geometrical quantities to serve as i, or 
embed $N$ points in a uniformally random way in each geometry and then use the pairwise metric 
distances between the points to furbish a vectorial i.  
Or, one could use total volume, anisotropy parameter or a vector made out of these, or use curvature 
invariants such as maximal or average curvatures of a given 3-space (e.g. objects related to the Weyl 
tensor considered in \cite{Weyls} which are also perported measures of gravitational information, see 
Sec 5).      
Or, for nonhomogeneous GR, one could compute eigenvalues of an operator D associated with that geometry 
and extract a spectral measure i from these.\fn{The counterpart of 
this for RPM's is trivial, as the configuration space is not itself a space of geometries.}
For example, Matzner \cite{Matzner} considers the first eigenvalue of the divergence of the Killing 
operator.
%
%
Spectra, however, can be shared by different geometries -- the isospectral problem.
Another measure of inhomogeneity that could be used as an i would be an energy density contrast 
type quantity F[$\varepsilon/\langle\varepsilon\rangle]$ (for $\varepsilon$ the energy density 
distribution and $\langle \mbox{ } \rangle$ denoting average over some volume) such as 
$\varepsilon/\langle\varepsilon\rangle$ or 
\be
\left\langle
\frac{\varepsilon}{\langle\varepsilon\rangle}\mbox{log}
\left(
\frac{\varepsilon}{\langle\varepsilon\rangle}
\right) 
\right\rangle
\label{star} \mbox{ } , 
\ee
which particular functional form also has information content connotations (see Sec 6).

For each structure above, one can readily supply a notion of `within $\epsilon$ of' (contingent 
to what distance axioms the structure obeys).   
We are now in a position to be able to give examples of grainings: just combine each example above with 
the obvious corresponding `within $\epsilon$' notion to obtain notions of `almost equal moment of 
inertia', `almost aligned', `almost equal size', `almost equal anisotropy' or 
`almost homogeneous geometry'.

If one takes a set of subconfigurations that are close in space according to a suitable notion and 
then partition that set according to a suitable notion of closeness in configuration space, one has 
a theoretical framework for {\sl localized configurational records (LCR's)}.   
RPM's with their local particle clusters, and inhomogeneous perturbations about minisuperspace 
with their localized bumps, are two modelling situations with a good notion of LCR.


This Sec permits one to address four further question types.  
Two are generalizations of their previously introduced primed counterparts, so as 
to model the imperfection of observation.  

\noindent Be$1$), does $q_{\Delta}(P)$ have acceptable properties? 
This is now for a graining set P rather than for an individual instant p.    

\noindent Be$2$), if properties of $q_{\Delta}(P)$ are known, does this permit deduction of any 
properties of $q_{\Delta^{\prime}}(P)$ for $\Delta^{\prime}$ disjoint from $\Delta$?

\noindent The other two involve the $\fQ$ space of the theory or theories that the observations are 
perported to belong to.  

\noindent
BeS$1$) is: what is $\mbox{P}(q_{\Delta}(P))$ within the collection of subconfiguration spaces? 
 
\noindent
BeS$2$) is: what is P($q_{\Delta^{\prime}}(P)$ has properties ${\cal P}^{\prime} |  q_{\Delta}(P)$ has 
properties ${\cal P}$)?\fn{Here, 
$|$ denotes `given that', so this is a conditional probability.  
These 
last 2 questions require one to determine or postulate a measure on $\fQ$; 
both this Sec and the next provide possible structures for this.}  

\noindent Examples of such questions are: what is P(particles 1, 2, 3 are almost colinear)? 
What is P(space is almost flat)? 
What is P(space is almost isotropic)? 
What is P(space is almost homogeneous)?

\subsection{Further structure: configuration comparers and decorated instants}

The above single-configuration notion of closeness may not suffice for some purposes (whether in principle or 
through lack of mathematical structure leaving one bereft of theorems through which to make progress).  
Other notions of closeness on the collection may depend on a fuller notion of comparison {\sl between} 
instants, i.e. their joint consideration rather than a subsequent comparison of real numbers extracted 
from each individually.   
That may either be a means of judging which instants are similar or a means of 
judging which instants can evolve into each other along dynamical trajectories.  
Some criteria to determine which notion should be used are adherence to the axioms of distance, 
gauge or 3-Diffeomorphism invariance as suitable, and, for some applications, 
whether they can be applied to grand heaps.  See \cite{ARec2} for further discussion.

One way of providing comparers is to upgrade the previous subsection's normed spaces and geometries to inner 
product spaces, metric spaces and topological spaces \cite{FORD, Kendall}.  
In the case of inner products or metrics, ${\cal M}^{\Gamma\Delta}Q_{\Gamma}Q^{\prime}_{\Delta}$ then 
supplies a primitive comparer of unprimed and primed objects $Q_{\Gamma}$, $Q_{\Delta}^{\prime}$.

Also consider replacing $\fQ_{\Delta}$ by 
the tangent bundle $\fT(\fQ_{\Delta})$ (configuration-velocity space \cite{B94I}), or 
the unit tangent bundle $\fT_u(\fQ_{\Delta})$ (configuration-direction space), or 
the cotangent bundle $\fT^*(\fQ_{\Delta})$ (configuration-momentum space, which, 
if augmented by a symplectic structure, is phase space). 
Such notions indeed continue to exist for restricted configuration spaces in cases with constraints.  
This last feature involves quotienting operations, which can considerably complicate structure in 
practise.    
Envisage all these as {\it `heaps of decorated instants'}, $\fH$, which more general notion I use 
to supercede $\fQ$.

A common situation is not to compare configurations $Q_{\Gamma}$ and $Q_{\Delta}^{\prime}$ but rather to 
compare the corresponding velocities $\dot{Q}_{\Gamma}$ and $\dot{Q}_{\Delta}^{\prime}$, with the
${\cal M}^{\Gamma\Delta}$ employed being the kinetic metric (which is the inverse of the object in the 
momentum form of quadratic constraints such as the Hamiltonian constraint or the energy constraint).  


An example of comparer constructed along these lines is the Lagrangian  
$\fL: \fT(\mbox{G-bundle over } \fQ) \longrightarrow \mathbb{R}$ 
\be
\fL[Q_{\Delta}, g_{\Lambda}, \dot{Q}_{\Delta}, \dot{g}_{\Lambda}] = 2\sqrt{\fT\{\fU + \fE\}} \mbox{ } ,  
\label{Lag}
\ee
where, in this paper's examples, $\fU$ is minus the potential term $\fV(Q_{\Delta})$ and 
$\fT$ is the kinetic term 
\be
\fT[Q_{\Delta}, g_{\Lambda}, \dot{Q}_{\Delta}, \dot{g}_{\Lambda}] = 
\frac{1}{2}M^{\Gamma\Delta}(Q_{\Theta})\{\stackrel{\longrightarrow}{G}\dot{Q}_{\Gamma}\}
                                         \stackrel{\longrightarrow}{G}\dot{Q}_{\Delta}
\ee
for $\stackrel{\longrightarrow}{G}$ the action of the group G of redundant motions whose generators are 
parametrized by auxiliary variables $g_{\Lambda}$.  
Here, the dot denotes the derivative with respect to label-time, an overall time that is meaningless 
because the actions considered are invariant under label change (reparametrization).

%
This also exemplifies that one often corrects the $Q_{\Gamma}$ or $\dot{Q}_{\Gamma}$ with respect to a 
group G of transformations under which they are held to be physically unchanged.  
That involves the group action of G on the $Q_{\Gamma}$ or $\dot{Q}_{\Gamma}$.  
E.g. in the case of particle velocities $\dot{q}_{i\alpha}$, 
the infinitesimal action of the translations (generated by $a_{\alpha}$) is 
$\dot{q}_{i\alpha} \longrightarrow \mbox{ } \stackrel{\rightarrow}{T} \dot{q}_{i\alpha} = \dot{q}_{i\alpha} + 
\dot{a}_{\alpha}$, 
the infinitesimal action of the rotations (generated by $b_{\alpha}$) is 
$\dot{q}_{i\alpha} \longrightarrow \mbox{ } \stackrel{\rightarrow}{R} \dot{q}_{i\alpha} = \dot{q}_{i\alpha} + 
q_{i\alpha} \cr \dot{b}_{\alpha}$   
and the infinitesimal action of the dilations (generated by c) is 
$\dot{q}_{i\alpha} \longrightarrow \mbox{ } \stackrel{\rightarrow}{D} \dot{q}_{i\alpha} = \dot{q}_{i\alpha} + 
\dot{c} q_{i\alpha}$ .  
E.g. in the case of 3-metric velocities $\dot{h}_{\mu\nu}$, 
the infinitesimal action of the 3-diffeomorphisms (generated by $s_{\mu}$) is 
$\dot{h}_{\mu\nu} \longrightarrow \mbox{ } \stackrel{\rightarrow}{\mbox{Diff}} \dot{h}_{\mu\nu} = 
\dot{h}_{\mu\nu} - \pounds_{\dot{s}}h_{\mu\nu}$.   
One furthermore often then minimizes with respect to the group generator, viewed as an arbitrary frame  
`shuffling auxiliary'.  
This ensures the physical requirement of G-invariance (i.e. gauge invariance, including 3-diffeomorphism 
invariance in geometrodynamics).

Then one has, for example, the following comparers.   
1) The Kendall-type \cite{Kendall} comparer is 
\be
\stackrel{\mbox{\scriptsize min}}{\mbox{\scriptsize $g \in$ G}} 
{\cal M}^{\Gamma\Delta}{\cal Q}_{\Gamma}\stackrel{\rightarrow}{G} {\cal Q}^{\prime}_{\Delta} 
\mbox{ } 
\ee  
for $\stackrel{\rightarrow}{G}$ the finite group action.  
2) Construct 
\be
{\cal M}^{\Gamma\Delta}\{\stackrel{\rightarrow}{G} \dot{Q}_{\Gamma}\}
                         \stackrel{\rightarrow}{G} \dot{Q}_{\Delta}
\ee
for $\stackrel{\rightarrow}{G}$ the infinitesimal group action.  
Then weight by $\fU + \fE$, square-root to obtain (\ref{Lag}), and then integrate with 
respect to spatial extent if required and with respect to label time so as to produce the corresponding 
action.  
Variation of this ensures G-independence.  
Examples of actions of this form (`thin sandwiches') are the Jacobi action \cite{Lanczos} for mechanics, 
Barbour--Bertotti type actions for RPM \cite{BB82}, and the Baierlein--Sharp--Wheeler \cite{BSW} 
(see also \cite{BB82, B94I, B94II, EOT, RWR, ABFKO}) type actions for geometrodynamics on a fixed 
compact without boundary topology.  
The variational procedure then entails minimization with respect to $g_{\Lambda}$.\fn{One
can envisage thin sandwiches as limits of 2-configuration `thick sandwich' expressions.   
Then variation with respect to $g_{\Lambda}$ gives a constraint; viewed in Lagrangian picture, this is 
an equation in $\dot{g}_{\Lambda}$ -- a thin sandwich equation.  
That this can be solved for $\dot{g}_{\Lambda}$ is the thin sandwich conjecture.  
If it can be solved for a particular problem, one can then use this equation to pass to a reduced action 
(see below).}  
%
One could also weight by $1/\{\fU + \fE\}$ and square-root.  
This gives Leibniz--Mach--Barbour timefunctions (c.f. \cite{B94I, SemiclI, SemiclII}).  
One could also not weight, giving a `kinematical' rather than `dynamical' comparer.  
Another variant is the DeWitt measure of distance: let one $\dot{h}_{\alpha\beta}$ 
and 1 of the 2 metrics in each factor of the DeWitt supermetric be with respect to primed coordinates, 
integrate with respect to both primed and unprimed space, and {\sl then} square-root.   
One then obtains a semi-Riemannian metric functional \cite{DeWitt70} 
(in the sense of `Finslerian metric function').  
For minisuperspace, integrations with respect to space are trivial and shuffling with respect to 
3-diffeomorphisms is trivial.  
What survives as a nontrivial construct is a combined measure of volume and anisotropy difference, 
that is decomposable into separate volume and anisotropy comparers.  
In full and perturbative inhomogenous geometrodynamics, one can likewise decompose combined measures of 
local size and shape (some of which resemble techniques used in \cite{ABFO, ABFKO}) into separate 
comparers \cite{ARec2}. 
In each case, the individual rather than combined comparers are better-behaved as notions of distance 
\cite{ARec2}).  
%

Comparers along the lines of 1) and 2) are universal, insofar as they apply both to RPM's and to GR.  
One difference however is that the GR version has an indefinite inner product which does not confer 
good distance properties in contrast to the positive definite one in mechanics.  
Thus one might need different tools in each case, so the non-universality of the below tools should not 
be held too much against them. 
On the other hand, one might get round this by using the {\sl shape part} of the GR inner product, 
which is itself positive definite \cite{ARec2}.  
 
3) In the case of 3-metrics, another comparer whose Diff-independence is assured by a similar method to 
the above is Gromov's distance between Riemannian spaces \cite{Gromov}.

\mbox{ }

Instead of using highly redundant variables alongside gauge auxiliaries and a shuffling procedure, one 
could work with reduced gauge-invariant configurations $Q_{\Omega}$, for $\Omega$ a smaller indexing set 
than $\Delta$, and a Lagrangian $\widetilde{\fL}: \widetilde{\fT}(\fQ_{\Omega}) \longrightarrow 
\mathbb{R}$ constructed from these, 
\be
\widetilde{\fL}[Q_{\Omega}, \dot{Q}_{\Omega}] = 
2\sqrt{        \widetilde{\fT}    \{    \widetilde{\fU} + \widetilde{\fE}    \}         }
\ee 
for $\widetilde{\fT}[Q_{\Omega}, \dot{Q}_{\Omega}]$ a suitable, `more twisted' kinetic term.  
The trouble with this approach is that one seldom has the luxury of explicit gauge-invariant variables 
being available.  
One case in which they are available is the 2d RPM of pure shape \cite{FORD}.
Building a mechanics from the natural objects in Kendall's study of the configuration space, or, 
equivalently, reducing Barbour's scale-invariant RPM (which amounts to solving the thin sandwich in this 
case) permits the explicit metric on the reduced configuration space to be evaluated in this case.  
It is the Fubini--Study metric (\ref{FS}), whereupon this case's `more twisted' kinetic term is  
\be
\widetilde{\fT}({\cal Z}_{\sfc}, \dot{\cal Z}_{\sfc}) = 
\frac{1}{2}\frac{\{1 + \sum_{\sfc}|{\cal Z}_{\sfc}|^2\} \sum_{\sfc}|\dot{{\cal Z}}_{\sfc}|^2 - 
                 |\sum_{\sfc} {\cal Z}_{\sfc} \dot{\overline{{\cal Z}}}_{\sfc}|^2}
              {\{1 + \sum_{\sfc}|{\cal Z}_{\sfc}|^2\}^2}  \mbox{ } 
\label{FSkinterm} \mbox{ } .  
\ee
One can then use notion of distance $D$ associated with this metric as a measure of distance between 
shapes in 2d space.\fn{Kendall also provides tools for higher-d spaces 
%
%
but these are a lot harder to use and somewhat less developed.}

Alternatively, one could work with (more widely available) secondary quantities that are guaranteed to 
have the suitable invariances.  
E.g. Seriu's spectral measure \cite{Seriu} 
\be
D_{\mbox{\scriptsize Seriu}}((\Sigma, h_{\mu\nu}), (\Sigma^{\prime}, h_{\mu\nu}^{\prime})|) = 
\frac{1}{2}\sum_{k = 1}^{\infty}\mbox{log}\left(\frac{1}{2} 
\left\{
\sqrt{\frac{\lambda_k}{\lambda_k^{\prime}}} + \sqrt{\frac{\lambda_k^{\prime}}{\lambda_k}}
\right\}\right) 
\mbox{ } , 
\ee
where $\{\lambda_k,\mbox{ $k$ = 0 to } \infty\}$ and $\{\lambda_k^{\prime}, \mbox{ $k$ = 0 to } \infty\}$ 
are the whole spectrum of the Laplace operator on the primed and unprimed manifolds respectively.  
This (and spectral measures in general) allow for comparison between geometries with different 
topology.  This particular form has quantum--cosmological significance but unfortunately fails 
to obey the triangle inequality as well as having an isospectral problem.

  
If there's a sense of more than one instant, there is one becoming question type per question 
type above, by the construction P(if $Q_{\Delta}(p)$ has properties ${\cal P}$ then it becomes 
$Q^{\prime}_{\Delta}(p^{\prime})$ with properties ${\cal P}^{\prime}$).  
I denote each such question type as above but with `Become' rather than `Be'.

\subsection{If there were a notion of time}

Then yet further question types would emerge.  
One source of novelty corresponds to each of the non-statespace questions, involving 
each instant being furthermore prescribed to be at a time, which I denote by appending a T.  
The new Be's are questions of `being at a particular time',     
while the new Become's are of the form `X at time 1 becomes Y at time 2'.  
For questions concerning heaps there is one extra ambiguity: 
`at any time' now makes sense in addition to `at a particular time'.   
This means that for each BeS question there are two BeST questions (denoted a, b), 
and for each BecomeS question there are four BecomeST questions (denoted a, b, c, d).  
Thus (c.f. Fig 1), 32 question-types have been uncovered.  

\mbox{ }

\noindent 
[{\bf Fig 1}.  The various question types and which moves remove some of them as separate entities.  
Single-subconfiguration questions (no S's) are at most consistency checks, while S-questions 
inter-relate observations and are thereby {\sl testable} if there is another LCR to bring into the 
picture.]

\subsection{Further analysis of question-types and of time}

First note {\bf Suppression 1}: the 8 primed questions are clearly but subcases of their more realistic unprimed counterparts.

Next note that the previous subsection crucially does not say what time is.  
Ordinary classical physics has an easy disculpation: there is an external time belonging to the real 
numbers, so that each $\fH$ is augmented to an extended heap space $\fH
\times \mathbb{R}$.
One key lesson from GR, however, is that there is no such external time.    
Stationary spacetimes (including the Minkowski spacetime of SR) do possess a timelike Killing vector, 
permitting a close analogue of external time to be used, but the generic GR solution permits no such 
construction.
The generic solution of GR has a vast family of coordinate timefunctions, none of which has a privileged 
status unlike that associated with the timelike Killing vector of a stationary spacetime.    
Questions along the lines of those above which involve time need thus specify {\sl which} time.  
Using `just any' time comes with the multiple choice and functional evolution \cite{K92, Isham93} 
subaspects of the POT -- there is a tendancy for this ambiguity to lead to inequivalent physics 
at the quantum level.

Another way of latching onto some aspects of the above key lesson, which moreover can already be 
modelled at the level of nonrelativistic but temporally-relational mechanical models, is that `being, 
at a time $t_0$' is {\sl by itself} meaningless if one's theory is time label reparametrization 
invariant.

Alternatives that render particular times, whether uniquely or in families up to frame embedding variables, 
meaningful are specific internal, emergent or apparent time approaches.
Therein, time is but a property that can be read off the (decorated sub)configuration.  
E.g. York's internal time \cite{York72, Kuchar81, K92, Isham93} can be thought of in this way, 
as can the notion of time in the Page--Wootters approach \cite{CPI}.  
Thereby one has 

\noindent {\bf Subsumption 2}: all question types involving a T are turned into the corresponding 
question types without one.\fn{It is not clear which as in this setting one can have in principle 
different configurations take the same time value (e.g. through lying on different paths of motion).}  
Perhaps this property concerns a particular subconfiguration lying entirely within the 
$H_{\Delta}(P)$ in question (`a clock within the subsystem').     
Perhaps it concerns a subconfiguration lying within $H_{\Upsilon}(P)$ but entirely outside 
$H_{\Delta}(P)$ (`clock within the environment'/`background clock').
Though perhaps a clock subsystem could be part-interior and part-exterior to the
$\fH_{\Delta}$ in question.  
Indeed, one could have a universe-time to which all parts of the configuration contribute 
rather than a clock {\sl sub}system.

\noindent {\bf Subsumption 3}: Each BecomeST b, c pair becomes a single question type if there is time 
reversal invariance.  

\noindent {\bf Subsumption 4}: If the time used is globally defined on $\fH$, BeSTb questions and 
BecomeSTd questions are redundant.     
This can in any case be attained by considering restricted $\fH$ defined so that this is so.   
(Whether that excludes interesting physics is then pertinent).  

\noindent At this stage, one is left with a $2\times 2 \times 2$ grid (Fig 1).

\noindent {\bf Subsumption 5} has been suggested by Page (e.g \cite{PAOT}) and also to some extent 
Barbour \cite{EOT}.   
It consists in supplanting all becoming questions by more operationally accurate being questions as 
follows.    
It is not the past instant that is involved, but rather this appearing as a memory/subrecord in 
the present instant, alongside the subsystem itself.   
Thus this is in fact a correlation within the one instant.  
In this scheme, one does not have a sequence 
of events but rather one present event that contains memories or other evidence of `other events'.\fn{As 
an illustrative sketch, one can imagine a configuration in which the LCR actually under study is the 
na\"{i}ve LCR plus the observer next to it, whose memory includes a subconfiguration which encodes 
himself peering at the LCR `at an earlier time' and a subconfiguration in which he has this first 
memory and a prediction `derived from it'.}

If subsumption 5 is adopted, the remaining question types are Be2 about how likely a correlation between 
two subsystems within the one grained subinstant is, theory-observation question type BeS2 about how 
likely an instant is within a statespace, and two `consistency' question types Be1 and BeS1 about 
properties of a subinstant.

If subsumption 5 is not adopted (or not adoptable in practise), there are additionally four corresponding 
types of becoming questions. 
Reasons why subsumption 5 might not be adopted, or might not be a complete catch-all of what one would 
like to be explained include 
I) impracticality: studying a subsystem S now involves studying a larger subsystem containing multiple 
imprints of S.
Models involving memories would be particularly difficult to handle (see footnote 20). 
II) If one wants a scheme that can explain the Arrow of Time, then Page's scheme looks to be 
unsatisfactory.  
While single instants such as that in footnote 20 could be used to simulate the scientific process as 
regards `becoming questions', it is noteworthy that these single instants correspond to the {\sl latest} 
stage of the investigation (in the `becoming' interpretation), while `earlier instants' will not have 
this complete information.  
III) Additionally, important aspects of the scientific enterprise look to be incomplete in this approach 
-- in interpreting present correlations, one is in difficulty if one cannot affirm that one did in fact prime the measuring 
apparatus would appear to retain its importance.    
I.e. as well as the `last instant' playing an important role in the interpretation, initial conditions 
implicit in the `first instant' also look to play a role (see also \cite{Hartle93, H99}).

\subsection{Addressing each question at the quantum level and beyond?}

At the classical level, one could either take certainty to be a subcase of probability, or note that 
even classically it is probabilities that are relevant in practise -- e.g. due to limits on precision of 
observations.    
2) A notion of $\mbox{P(trajectory goes through a subregion $\Delta$}$ 
for each space $\fH$) is then required (see e.g. \cite{HT}).
This is particularly common in the literature in the case in which $\fH = \fP$, phase space.

Then if one canonically-quantizes, the Hamiltonian ${\cal H}$ provides a TISE
\be
\hat{\cal H}\Psi = 0 
\ee
such as (\ref{TISEO}) for ordinary mechanics, (\ref{TISE}) for RPM's in redundant variables, (\ref{TISE2d}) 
for 2d RPM's in reduced variables or the WDE (\ref{WDE}) in GR.

The question about angular momentum for the RPM given a subsystem's angular momentum 
has an immediate quantum counterpart by employing the QM notion of angular momentum.  
This kind of question involves eigenvalues of operators, and so can be addressed as in familiar QM.

The colinear, almost flat, almost isotropic, almost homogeneous questions, since they are configurational 
questions, have obvious counterparts in configuration--representation QM.  
These questions concern pieces of the configuration space, in which sense they lie outside the usual 
domain of QM.  
The \NSI and the \CPI are two interpretations outside or beyond conventional QM formalism suggested 
to answer such questions.     
The \NSI serves to address the BeS1 version of this paragraph's questions,\fn{Another 
interesting question of this type still being debated in the literature is what is P(Inflation), for 
which \cite{HP88, InflTool} provide tools.} 
e.g. \cite{HP86} use the natural measure associated with the supermetric to address the 
almost flatness of the universe.  
The \NSI concerns only questions of relative probability because of non-normalizability issues.

On the other hand, the \CPI is one for addressing Be2 or BeS2 questions such as 
P(Particles 1, 2, 3 are almost aligned $|$ particles 4, 5, 6 are almost aligned) 
[all within a given instantaneous configuration].  
I also comment that some of Kucha\v{r}'s criticisms \cite{K92, Kuchar99} of these approaches can be 
interpreted as not accepting ab initio a separate `being' position rather than being conceptual or 
technical problems once one has adapted such a position.

\section{Are records typically useful?}

Records Theory requires A) for subconfigurations to be capable of holding enough information to address 
whatever issues are under investigation.  
Let us approach this using Information Theory \cite{Preskill}.  
Information is (more or less) negentropy, so a starting classical notion is the Boltzmann-like\fn{I 
choose units such that Boltzmann's constant is 1.}  
\be
I_{\mbox{\scriptsize Boltzmann}} = - \mbox{log}W
\label{Boltz}
\ee
for $W$ the number of microstates, evaluated combinatorially in the discrete case or taken to be 
proportional to the phase space volume in the continuous case.  
One could furthermore use such as Shannon information, 
\be
I_{\mbox{\scriptsize Shannon}}(p_x) = \sum_x p_x\mbox{log}p_x
\ee 
for $p_x$ a discrete probability distribution for the records, or  
\be
I_{\mbox{\scriptsize Shannon}}[\sigma] = \int \d\Omega \sigma\mbox{log}\sigma
\label{ShaCont}
\ee    
for $\sigma$ a continuous probability distribution.  
If one is considering records at the quantum level, then one could instead use such as von Neumann 
information, 
\be
I_{\mbox{\scriptsize von Neumann}}[\rho] = \mbox{Tr}(\rho \mbox{log}\rho)
\ee 
for $\rho$ the density matrix of the quantum system.
The Shannon and von Neumann notions are both based on the nlogn function, (which is 
{\sl the} positive continuous function consistent with regraining and has many further useful properties 
\cite{Wehrl}), 
and these notions furthermore tie in well with each other as regards classical--QM correspondence 
(see e.g. \cite{Wehrl}).  
Furthermore the von Neumann notion does survive the transition to relativistic QM, and that to QFT modulo  
a short-distance cutoff \cite{CC04, Wald}.  
As regards GR, it has been used in the context of black holes (see e.g. \cite{Wald}).     
One contention in interpreting (0) at the general level required for developing a POT strategy is that 
information is minus entropy and classical (never mind quantum) 
gravitational entropy is a concept that is not well understood or quantified for general spacetimes 
\cite{Weyls, Smolin, Brandenberger, RA, HBM, Wald}.  
Quantum gravity may well have an information notion 
\be
I[\rho_{\mbox{\scriptsize QGrav}}] = 
\mbox{Tr}\rho_{\mbox{\scriptsize QGrav}}\mbox{log}\rho_{\mbox{\scriptsize QGrav}} 
\mbox{ } , 
\ee
but either the quantum-gravitational density matrix is an unknown object since the  
underlying microstates are unknown, or, alternatively, one would need to provide an extra procedure for 
obtaining this, such as how to solve and interpret the WDE, which would be fraught with numerous further 
technical and conceptual problems.   
Rather than a notion of gravitational information that is completely general, a notion of entropy 
suitable for approximate classical and quantum cosmologies may suffice for the present study. 
Quite a lot of candidate objects of this kind have been proposed.  
However, it is unclear how some of these would arise from the above fundamental picture, while 
for others it is not clear that the candidate does in fact possess properties that make it a bona fide 
entropy.  
Monotonicity is one often-mentioned property (with which gravitational information candidates based on 
the Weyl tensor \cite{Weyls} have run into problems), 
while information/entropy is characterized by a number of further properties \cite{Wehrl} that it is not 
clear that the gravitational candidates have been screened for.  
Examples of cosmologically relevant information notions proposed to date that are manifestly related to 
conventional notions of information are Rothman--Anninos' \cite{RA} use of the continuous form of 
(\ref{Boltz}) and Brandenberger et al.'s \cite{Brandenberger} continuous version of (\ref{ShaCont}). 
Also, Hosoya--Buchert--Morita \cite{HBM} use 
\be
I_{\mbox{\scriptsize HBM}}[\varepsilon] = 
\int\d\Omega \varepsilon \mbox{log}\left(\frac{\varepsilon}{\langle\varepsilon\rangle}\right) 
\mbox{ } \mbox{ and } \mbox{ } 
I^{\prime}_{\mbox{\scriptsize HBM}}[\varepsilon] =  
\left\langle\varepsilon \mbox{log}\left(\frac{\varepsilon}{\langle\varepsilon\rangle}\right)\right\rangle
\label{KLspec} \mbox{ } , 
\ee
the first of which is a relative information type quantity (see Sec 6); the second is related to the 
first by a factor of 1/Volume = 1/$\int\d\Omega$, while, if one is to have a quantity to serve as a 
negentropy, I argue that one should use a probability distribution input 
$f = \varepsilon/\int\varepsilon \d\Omega$ (in the sense that $\int f \d\Omega  = 1$) rather than 
$\varepsilon$, whereby I obtain my alternative measure (\ref{star}), related to the first form 
above by a factor of 1/$\int \varepsilon \d\Omega$.  
The differences between these quantities amount to different normalizations in comparing inhomogeneous 
spaces; which quantity it is best to use depends on the exact context.  
Particle mechanics toy models of these objects also exist, based on number density rather than energy 
density and using discrete rather than continuous mathematics \cite{ARec2}.

B) However, whether there is a pattern in a record or collection of records (and whether 
that pattern is significant rather than random) involves more than just how much information is 
contained within.  
Two placings of the same pieces on a chessboard could be from a grandmasters' game and a frivolous 
random placing.  
%
Two similar-size samples of the same kind of sand could be a hoofprint and a random pattern due to the 
wind blowing.
What one requires is a general quantification of there being a pattern.  
There should be at least a partial link between this and information content, in that 
at least some complicated patterns require a minimum amount of information in order to be realized.   
Records theory is, intuitively, about drawing conclusions from similar patterns in different records.

Consider also the situation in which information in a curve or in a wave pulse that is  
detectable by/storeable in a detector in terms of approximands or modes.  
As regards localized useable information content per unit volume, considering the Joos--Zeh dust--CMB  
and $\alpha$-track--bubble chamber side by side suggests that most records in nature/one's model will be 
poor or diffuse.
For the Joos--Zeh \cite{JZ} example the `somewhere' is all over the place: 
``{\it in the vastness of cosmological space}".   
Detectors, such as the extension of Halliwell's 1-piece detector model (Sec 3.3, \cite{H99})
to a cluster, could happen to be tuned to pick up the harmonics that are principal contributors in the signal.  
In this way one can obtain a good approximation to a curve from relatively little information.  
E.g. compare the square wave with the almost-square wave that is comprised of the first 10 harmonics 
of the square wave.  
That is clearly specific information as opposed to information storage capacity in general.  
Likewise, a bubble chamber is attuned to seeing tracks, a detector will often only detect certain 
(expected) frequencies.  
Through such specialization, a record that `stands out' can be formed.    
One should thus investigate is quantitatively which of the $\alpha$-track and `dust grain' paradigms 
is more common.

C) Information can be lost from a record `after its formative event' -- the word ``stored" in (0) 
can also be problematic.     
Photos yellow with age and can be defaced or doctored,
while some characteristics of the microwave `background' radiation that we observe have in part been 
formed since last scattering by such as the integrated Sachs--Wolfe effect or foreground effects 
\cite{coscorr}.

\section{Simple models of correlations between records}

One concept of possible use is {\it mutual information}: this is a notion 
\be
M(A, B) = I(A) + I(B) - I(AB)
\ee
for AB the joint distribution of A and B for each of classical Shannon or 
QM von Neumann information \cite{Preskill}.  
This is a quantity of the relative information type \cite{KL},    
\be
I_{\mbox{\scriptsize relative}}[p, q]         = \sum_x p_x\mbox{log}\left(\frac{p_x}{q_x}\right) 
\mbox{ } \mbox{ (discrete case) ,} \mbox{ } \mbox{ }
I_{\mbox{\scriptsize relative}}[\sigma, \tau] = \int \d\Omega \sigma\mbox{log}\left(\frac{\sigma}{\tau}\right) 
\mbox{ } \mbox{ (continuous case) ,} 
\label{relen} 
\ee
(the first form of (\ref{KLspec}) is also a special case of the continuous case of this in which the role 
of the second distribution is played by the average of the first).  
Tsallis relative information \cite{Tsallis} is a more general such notion which includes both 
(\ref{relen}) and Shannon information as special cases.    
The QM counterpart of relative information is \cite{NC}
\be
I_{\mbox{\scriptsize relative}}[\rho_1, \rho_2] = \mbox{Tr}(\rho_1\{ \mbox{log}\rho_1 - \mbox{log}\rho_2\})
\ee
There is also a QM analogue of mutual information (and a further closely-related notion, 
{\it Holevo information}  \cite{Preskill}, and a quantum analogue of Tsallis information 
\cite{Tsallis}).      
Such may be of use in the quantum cosmological counterpart of Hosoya--Buchert--Morita type 
inhomogeneity measures \cite{ARec2}.

It is not clear that these notions cover all possibilities as regards patterns.  
Two records could be part of a discernible common pattern even if their constituent information is 
entirely different, e.g. the pattern to spot on two chessboards could be interprotection, 
manifest between rooks on one board and between knights on the other.

Next consider the notion of measures of correlation.   
As a first example, I consider approximate collinearity as an in some ways simple kind of correlation.  
There are various types of collinearity.  
The most elementary involves seeking for a single line of best fit for data in $\mathbb{R}^2$ and 
assessing whether this is significant for that data (e.g. Spearman's test).  
This is not always appropriate.  
E.g. for the observed distribution of monoliths
or quasars, one would rather look for whether there are more quasi-alignments than is probable between 
each individual grouping of several objects. 
One would use e.g. the $\epsilon$-bluntness notion on all triples of points or other approaches by 
Kendall.\fn{Kendall \cite{Kendall84, Kendall} provides probability measure (Borel measure) structure on 
e.g. S(N, 2) which is the reduced configuration space for Barbour's scale-invariant RPM in 2d, and 
statistics that serve to test whether there is significant alignment of triples of points within a data 
set that lies within a polygonal or convex set.
Kendall also provides statistical methods to test more general propositions involving 
polygons formed by data points.}

As a second example, I consider the family of notions of correlator/n-point function that occur in 
cosmology, QFT and indeed QFT on a cosmological background.\fn{In cosmology, one has also searched for 
circles in the microwave background data as possible signatures of the large-scale shape of the universe 
\cite{circles}, while CMB analysis concerns further features of inhomogeneities that go beyond relative 
spatial positions such as the spreads and shapes of hot and cold spots.}       
I consider the 2-point function for simplicity (but the analysis readily generalizes to the n-point 
function).  
In classical cosmology, one works with the 2-point function for such as mass density or galaxy number 
density \cite{coscorr}, 
\be
T(\underline{r}) = 
\frac{\langle\sigma(\underline{r}^{\prime})\sigma(\underline{r}^{\prime} - \underline{r})\rangle - 
      \langle\sigma(\underline{r})\rangle  \langle\sigma(\underline{r}^{\prime} - \underline{r})\rangle}
{\langle \sigma(\underline{r})\rangle  \langle \sigma(\underline{r}^{\prime} - \underline{r})\rangle}   
\ee
(or some integrated, angular or Fourier-transformed 
version of this, the Fourier transform being the well-known {\it power spectrum} quantity).   
QM has an extra kind of correlations that classical theory does not have, due to entanglement.   
Wigner functions are also available as tools at the quantum level (see e.g. \cite{HT}).  
Ordinary (Minkowski spacetime) QFT has n-point functions of the same kind \cite{qftcorr}, where 
$\langle \mbox{ } \rangle$ now includes inserting the ground-state wavefunction at each end.      
This notion carries over to n-point functions for a simple `Mukhanov variables QFT' picture of quantum 
cosmology \cite{qftcoscorr}.    
The Zalaletdinov correlation tensor (a comparer) \cite{Zala} applies to the more general inhomogeneous 
setting.  
On the one hand, these objects are higher moments, which indeed in general would be expected to carry 
more detail of structure than the above relative information quantities, but on the other 
hand it is not clear whether they can catch particular subtle details such as quasialignments between 
all triplets of included objects.   
\cite{Giddings} has a useful treatment of correlators for quantum cosmology.

\section{Conclusion}    

\subsection{A simple outline of Records Theory}    

Timeless Records Theory concerns subconfigurations of a single instant.     
These subconfigurations (`the records') are to be localized and only approximately known 
in order to be useable for study and in correspondence with what one meets in practise.  
The notion of `approximately known' is underpinned more precisely by a notion of closeness on the 
configuration space and a notion of graining.  
`Configuration space' above may be replaced by `decorated' versions such as 
configuration-velocity, configuration-direction or configuration-momentum (phase) space.  
For records to be useable, they need to contain enough of the right kind of information (such as a 
common pattern, shared information, or correlations) to furbish a deductive process.  
These features considered in this paper are useful `groundwork' for Records Theory, but addressing some 
questions of interest in Quantum Cosmology would require further ideas and structures to be built up.  
The paper has reviewed a number of notions of closeness, information content, information comparison 
and correlation that could provide Records Theory with formal mathematical machinery, but exactly which 
notions should be adopted is left open to be tested out in ongoing investigations \cite{ARec2} with 
various explicit toy models.   
Some economy of structure may be possible, in that I have shown how some versions of various different 
notions can be based on a common mathematical structure.

Records theory can manifestly cope with questions about `being'.  
As regards the other kinds of questions that one may have in dynamics/science/history, questions 
involving `at a particular time' can be supplanted by using (part of) the configuration to give meaning 
to and determine that notion of time.  
Questions concerning `becoming' can in principle be rephrased as question of `being', since the `latest' 
relevant instant will contain (not necessarily useable) memory-like records of the `previous' relevant 
instants, which would then be `reconstructed'.  
Records would furthermore need to fulfill `semblance of dynamics' in order to explain the world we see.  
How this would come about goes beyond the scope of what this paper covers in detail; see Sec 7.3 
for brief discussion.

\subsection{How motivation for records should be presented}  

It is the following four arguments that motivate Records Theory.   
1) The Problem of Time (POT) in Quantum Gravity is an incompatibility between the roles played by `time' 
in GR and in QM.  
One conceptually clear way of dealing with this problem is to recast both GR and QM in a timeless mold.  
[While the Wheeler--DeWitt equation's timelessness might specifically prompt some physicists toward 
Timeless Records Theory, this equation has numerous technical problems and may not be trustworthy.   
Nor should one turn to Timeless Records Theory due to earlier detailed documentation of problems 
with the other POT approaches, but rather judge it due to its own merits and shortcomings (Sec 7.3).] 
2) One can in principle treat all of change, processes, dynamics, history and the scientific enterprise 
in these timeless terms (but it is difficult in practise, and, at least by the scheme provided in this 
paper, this does not suffice to explain the Arrow of Time).  
3) Records Theory is of potential use in the foundations of Quantum Cosmology 
(which is what Inflationary Theory is to rest on).  
4) Records Theory is (alongside Histories Theory) a universal scheme in that all types of theory or  
system admit a such (but it may not always be a profitable approach to take, due to the above useable 
and useful criteria, which does some damage to the claim that records are more fundamental than 
histories).

\subsection{Comment on less simple features of records and Records Theory}

What does it mean for a record to encode a semblance of dynamics?  
%
%
How does a record achieve this encodement? 
Are subconfigurations that encode this generic?  
Let us suppose that this is actually a special rather than generic feature for a subconfiguration to 
have.  
This would be the case if the dust grain--CMB photon paradigm is more typical than the 
$\alpha$-particle--bubble chamber one.  
Then one would have the problem of explaining why the universe around us nevertheless contains a 
noticeable portion of noticeably history-encoding records, i.e. one would then need a {\it selection 
principle}.

Barbour suggests a selection principle based on the following layers \cite{EOT}. 
1) There are some distinctive places in the configuration space.
2) The wavefunction of the universe peaks around these places, making them probable.   
3) These parts of the configuration space contain records that bear a semblance of dynamics 
(`time capsules').
[Following my arguments in this paper, this should be rephrased in terms of subconfigurations.]  
The following doubts may be cast on this scheme. 
Firstly, at least his earlier arguments \cite{B94I, EOT}, he suggests the wedge-shape of his 
representation of Triangle Land could play a role, but this is a representation-dependent rather than 
irreducible feature of Triangle Land, absent e.g. in the spherical representation. 
Secondly, in modelling small atoms 
and cosmic strings \cite{KS91}, the effect on the dynamics of representation-invariant features such as 
stratifications, while non-negligible, only impart a small distortion on the wavefunction.
Thirdly, Barbour supplies no concrete mathematical model evidence for there being any correlation 
between subconfigurations being time capsules and their being near a distinctive feature of 
configuration space such as a change of stratum or a point of great uniformity.

Semiclassicality might either explain or supplant Barbour's selection principle
(it is regions in which the WKB ansatz happens to apply that have a semblance of time?)
Barbour agrees to some extent with this.  
Additionally, there are two a priori unrelated selection principles in the literature, which 
could be viewed either as competitors or as features that Barbour's scheme should be checked to 
be able to account for.


I) In the study of ({\it Branching Processes}), one learns that Barbour's `how probable a subconfiguration 
is' can depend strongly on the precise extent of the list of its contents.    
As an example (closely paralleling Reichenbach \cite{Reichenbach}),  
suppose we see two patches of sand exhibiting hoof-shaped cavities. 
Then past interactions of these two patches of sand with a third presently unseen subsystem -- a horse  
that has subsequently become quasi-isolated from the two patches -- is clearly capable of rendering the 
individually improbable (low entropy and hence high information) configurations of each patch of sand 
collectively probable (high entropy, low information) for the many sand patches--horse subsystem.  
This still does not explain why useful records appear to be common in nature: a separate argument 
would be needed to account for why Branching Processes are common.    


II) A selection principle for records might also arise as a knock-on from there being a `consistency of 
histories within a graining' selection principle in Histories Theory. 
Perhaps not all record grainings correspond to projections of consistent history grainings.   
That the notion of inconsistent record grainings may not be absurd may follow from considering what one 
would obtain if one projected grainings in {\sl inconsistent} Histories Theory.  
Although, conceivably, selection could also be lost due to the projection, or be generated by the 
projection  --  if the consistent history involves filming and subsequently doctoring the tape, is that 
tape a consistent record?\fn{One  
should bear in mind, however, that Histories Theory has more structure than Records Theory.  
On the one hand, this might permit it to effectively model more situations, but on the other hand 
it is not clear whether all of History Theory's structure is operationally meaningful.   
Thus Records Theory's greater sparseness may serve as a useful probe of Histories Theory.}

Investigation of Barbour's selection principle could be done using a relational particle model 
complicated enough to have a distinctive curved configuration space geometry, such as the 
3-particle 2d model or the 4-particle 1d scalefree one.
This investigation would probably also benefit from comparison with parallel Semiclassical Approach and 
Histories Theory calculations.   
Barbour favours his `being' perspective so as to be open to the possibility of explaining the Arrow of 
Time, while I), II), Castagnino's scheme \cite{CastAsym} (which builds in a time asymmetry in the choice 
of admitted solutions), and Page's scheme (which is subject to the difficulties pointed out in Sec 4.4) 
are not open to such a possibility.

While specific examples will serve to strengthen our understanding of whether selection principles are 
necessary, how the semblance of dynamics is to be recovered and whether this permits an emergent rather 
than presupposed Arrow of Time, the context in which to assess whether Records Theory is a Problem of 
Time resolution is the fully generic one.  
Limitations of Records Theory in this regard due to features exposed in this paper are as follows 
(see also \cite{Giddings}).    
Records are ``somewhere in the universe that information is stored when histories decohere".     
But a suitable notion of localization in space and in configuration space may be hard to come by and/or 
to use for quantum gravity in general -- `where' particular records are can be problematic to quantify, 
and the records can be problematic to access and use too, since the relevant information may be 
`all over the place'.  
Also, `information' is problematic both as it may be of too poor a quality to reconstruct the history 
and because a suitably general notion of information is missing from our current understanding of 
classical gravity, never mind quantum gravity with its unknown microstates (mechanical toy models are 
useful in not having this last obstruction).  
Finally, the further Records Theory notions of significant correlation patterns and how one is to deduce 
dynamics/history from them looks to be a difficult and unexplored area even in simpler contexts than 
gravitation.  

\mbox{ } 

\noindent{\bf{\Large Acknowledgments}}

\mbox{ }

\noindent I thank: Dr Julian Barbour and Professor Don Page for many discussions on time, histories and 
records.  
Dr Luca Bombelli, Dr Jeremy Butterfield, Dr Fay Dowker, Professor Gary Gibbons, Professor Jonathan 
Halliwell and Professor Bill Wootters for various discussions, comments and references.   
Dr Julian Barbour and Dr Jeremy Butterfield for comments on earlier versions of the manuscript.     
My wife Claire for her support.   
The Killam Foundation for funding me in 2005 and Peterhouse for funding me since.


\end{document}